# Evaluation of a Virtual Laboratory Platform in General Education on Quantum Information Science

Hongbin Song

General Education Division, The Chinese University of Hong Kong, Shenzhen, China

Email: songhongbin@cuhk.edu.cn

*Abstract*—Quantum information science and technology has been revolutionizing our daily life, which attracts the curiosity of young generations from diverse backgrounds. While it is quite challenging to teach and learn quantum information science for non-physics majors due to the abstract and counter intuitive nature of quantum mechanics. As an essential resource in quantum information science, quantum entanglement plays an important role in various quantum information systems. Therefore, it is crucial for students to grasp the unique properties of quantum entanglement. However, the counterintuitive nature makes it particularly challenging for undergraduates to comprehend this important phenomenon. To address these challenges, virtual laboratories have emerged as an effective solution. This paper presents the findings of pedagogical research on the efficacy of a virtual laboratory platform in general education courses on quantum information science. Specifically, a virtual laboratory activity based on the Bell test has been developed using a commercially available Quantum Optical Simulation Laboratory, QLab. The experiential activity is designed to help undergraduates from diverse academic disciplines understand the counterintuitive yet foundational concept of quantum entanglement. Qualitative and quantitative evaluations conducted over three academic years using carefully designed questionnaires indicated that the virtual laboratory enabled over 80% of students to grasp the complex concepts of quantum entanglement. These results demonstrate the effectiveness of the virtual laboratory in making abstract quantum concepts accessible and engaging, regardless of students' prior knowledge of advanced mathematics or technical skills. Despite certain limitations, such as the relatively small sample sizes in the last two semesters, this study offers valuable insights and a practical framework for addressing the challenges of teaching quantum information science in undergraduate curricula, particularly within general education courses designed for both science and non-science students.

## I. Introduction

Quantum mechanics, as a cornerstone of modern physics, uncovers the fundamental principles governing the microscopic world and has revolutionized our understanding of nature. Its applications have led to groundbreaking technologies such as lasers, semiconductors, and nuclear magnetic resonance imaging (NMRI) [1], profoundly shaping various aspects of daily life. Moreover, advancements in quantum information science and technology, including secure communication [2], quantum advantages over classical computers [3-5], and the detection of gravitational waves [6], are driving us toward a new information revolution. Recognizing the importance of these developments, the World Quantum Day was launched on April 14, 2021, to engage the general public in understanding and discussing quantum science and technology [7]. The first global celebration took place on April 14, 2022, marking a significant milestone in fostering awareness and enthusiasm for quantum advancements [7]. Building on this momentum, dozens of national scientific societies came together to commemorate 100 years of quantum mechanics. On June 7, 2024, the United Nations proclaimed 2025 as the International Year of Quantum Science and Technology (IYQ) [8]. This year-long global initiative aims to increase public awareness of the importance of quantum science and its applications, stating that it will "be observed through activities at all levels aimed at increasing public awareness of the importance of quantum science and applications" [8]. These initiatives reflect the growing global recognition of the transformative potential of quantum science and highlight the pressing need for expanded public engagement and education in this new era. However, most quantum mechanics curricula in higher education are currently designed for students majoring in physics or engineering. This creates an urgent need to develop courses accessible to all undergraduates, including those from both science and arts academic backgrounds. As we navigate this pivotal moment, addressing the pedagogical challenges of effectively teaching quantum mechanics and its interdisciplinary applications is becoming increasingly critical. Equipping learners with the perspectives and knowledge needed to understand and contribute to this rapidly evolving field is essential for fostering a well-prepared and informed generation capable of engaging with the transformative advancements of quantum science.

Since 2015, we have developed and implemented general education courses on quantum information science and technology at the Chinese University of Hong Kong, Shenzhen, to support the mission of integrating quantum information science into higher education. These courses are carefully designed to provide undergraduates from a wide range of academic disciplines with an accessible yet rigorous introduction to quantum mechanics and its applications [9]. By emphasizing both the historical development of quantum mechanics and its role in enabling the exploration of nature at its most fundamental level, the curriculum promotes a deep conceptual understanding that transcends disciplinary boundaries and fosters interdisciplinary learning.

In addition, the courses emphasize the transformative impact of quantum mechanics on modern technologies, its potential to drive future advancements in science and technology, and its broader societal implications. This integrative approach not only supports students in appreciating the relevance of quantum science in their daily lives but also aligns with the objectives of the International Year of Quantum Science (IYQ) [7, 9], which seeks to make quantum science more accessible and impactful across educational contexts.

The historic debate between Albert Einstein and Niels Bohr highlights their differing interpretations of the nature of reality, as reflected by the contrast between Einstein's views and the Copenhagen school's perspective [10]. The EPR paradox, proposed by Einstein and his colleagues in their 1935 paper, intensified this debate by arguing that quantum mechanics, as interpreted by the Copenhagen school, is either incomplete or involves what Einstein famously referred to as "spooky action at a distance" [11]. Despite Niels Bohr's defense of the Copenhagen interpretation, no definitive conclusion was reached at the time [12]. The debate remained unsettled until John Bell's seminal 1964 paper, which provided a theoretical framework—now known as Bell's Theorem—to experimentally test the validity of quantum mechanics [13]. Pioneering experiments conducted by John F. Clauser, Alain Aspect, and Anton Zeilinger provided empirical evidence that the quantum correlations between entangled particles violate Bell's inequality, thereby confirming the existence of "spooky action at a distance" and resolving the decades-long debate [14]. Their groundbreaking work not only addressed foundational questions in quantum mechanics but also laid the groundwork for quantum information science. This breakthrough paved the way for the development and experimental realization of various protocols based on quantum entanglement, such as quantum teleportation, which are essential for quantum communication, quantum computing, quantum metrology, and quantum sensing. Their contributions were recognized with the Nobel Prize in Physics in 2022.

Therefore, it is crucial for students to understand the origins of the Einstein-Bohr debate, the EPR paradox, and how the Bell tests ultimately resolved these questions. Despite its significance, the abstract and counterintuitive nature of quantum entanglement presents considerable challenges in teaching and learning, particularly at the undergraduate level. Grasping these concepts is especially difficult without direct observation of the phenomena through experiments. However, unlike the hands-on experiments typical of classical physics, conducting such complex quantum experiments in a real laboratory is impractical for undergraduates, especially those from diverse academic backgrounds, due to limitations in both equipment and experimental skills.

In this paper, we address the pedagogical challenges of teaching quantum information science to undergraduates from diverse academic backgrounds, including both science and arts. We introduce the Bell test experiment using a commercially available virtual laboratory platform, QLab from Anhui Qasky Quantum Technology Co. Ltd. This innovative platform simulates a 3-D physical laboratory environment equipped with various optical devices and instruments, enabling students to engage in hands-on experimentation with quantum entanglement- a conceptually challenging and abstract phenomenon. The virtual laboratory addresses key teaching barriers by overcoming the physical constraints of traditional equipment making it particularly valuable for institutions without access to specialized quantum optical labs. Additionally, it provides a safe, accessible, and interactive experiential learning environment that fosters creativeity and inclusivity, thereby supporting diverse learners in engaging meaningfully with complex quantum concepts. We evaluate the educational effectiveness of the virtual laboratory platform using both quantitative and qualitative methods, drawing on detailed student feedback to provide a comprehensive understanding of its impact on learning outcomes.

## II. Literature Review

Quantum mechanics is a foundational course for undergraduates majoring in physics or electrical engineering. Recently, many secondary schools in countries such as Australia, Canada (Ontario), Denmark, the UK, Finland, France, Germany (Baden-Württemberg), Portugal, and Spain have started offering quantum mechanics in their curricula [15-17]. However, as noted in Stadermann's work, only two out of 15 countries include quantum entanglement in their secondary school curriculum. This omission is primarily attributed to the conceptual and pedagogical challenges it poses to students at this level [15].

Studies have shown that even undergraduates from science and engineering backgrounds struggle to build mental models and visual representations of foundational quantum concepts [18, 19]. Teaching laboratories, as a quintessential method of practicing experiential learning thoery [20-22], play a crucial role in addressing these issues by providing students with opportunities to apply theoretical knowledge through carefully designed experiments tailored to their level of expertise or the specific topics covered in a course or program of study [23, 24]. To bridge these gaps, Lahoz Sanz reported a versatile and cost-effective system developed for undergraduates, enabling them to perform experiments such as measuring Bell inequalities and conducting quantum state tomography [25]. This system enhances accessibility for less specialized laboratories, allowing students to gain hands-on experience and build familiarity with core quantum physics concepts, highlighting the effectiveness of experientail learning. However, manipulating quantum optical systems remains particularly challenging for undergraduates, especially for non-physics or non-engineering majors in general education, due to the advanced skills required.

In addition to these challenges, laboratory safety has always been a significant pedagogical concern. Teaching environments, such as chemistry laboratories, often involve risks of explosions, while optical laboratories pose hazards from high-power lasers [26, 27]. These safety concerns can limit the extent to which students are allowed to interact directly with advanced experimental setups, thereby restricting hands-on learning opportunities. Moreover, maintaining laboratory instruments is difficult, as some devices are prone to damage from improper operation, which further discourages open student engagement. From a pedagogical perspective, these limitations hinder the creation of an effective and interactive learning environment, especially for courses on quantum information science. The complexity of quantum systems, combined with the need for precise handling of equipment, poses significant barriers to designing safe, accessible, and engaging experiments that align with students' learning needs and levels of expertise.

Virtual laboratories, which replicate physical laboratory environments or hands-on experimental activities in two- or three-dimensional digital formats, present a viable solution to the challenges previously highlighted [20, 25–27]. These platforms facilitate the exploration of scientific concepts and principles by enabling students to interact with virtual equipment and materials through interfaces such as keyboards or handheld controllers [31].

A key advantage of virtual laboratories is their ability to transcend physical and geographical limitations, offering accessibility anytime and anywhere. This flexibility enbales virtual laboratories to funcaion not only as a complement to traditional on-campus laboratory experiences but also, in certain cases, as a complete substitute, providing students with a convenient and adaptable approach to laboratory-based learning [27]. This feature is particularly beneficial during global crises, such as pandemics, or in remote learning contexts. Furthermore, the virtual nature of the required laboratory equipment significantly reduces the financial burden on institutions with limited resources to establish or maintain conventional laboratory facilities [30, 31]. Additionally, these platforms support collaborative learning by enabling students to work in teams, thereby fostering the development of critical teamwork and communication skills essential for both academic and professional success. Moreover, unlike immersive virtual reality systems, virtual laboratories minimize health and safety risks, providing a safer and more practical environment for experiential learning [20-22,28].

Consequently, virtual laboratories have gained widespread use in higher education, supporting both professional and non-professional study programs. They represent a modern advancement in STEM (science, technology, engineering and mathmetics) education by enhancing accessibility while minimizing the need for specialized laboratory infrastructure. At the same time, they preserve the immersive, hands-on learning experience that is a hallmark of traditional laboratory settings [30-36]. Pedersen et al. introduced a virtual learning environment, StudentResearcher, which integrates simulations, multiple-choice quizzes, video lectures, and gamification into the learning pathway for advanced university-level quantum mechanics [34]. A no-code online laboratory of an optical table, Virtual Lab by Quantum Flytrap, has been developed to present quantum phenomena interactively and intuitively [32]. With advancements in virtual reality technology, Müfit and colleagues developed an immersive virtual reality (IVR) platform incorporating cognitive conflict to support practical learning of quantum physics for undergraduate physics majors [36].

Reeves systematically reviewed and synthesized 25 peer-reviewed studies (2009–2019) on virtual laboratories (V-Labs) in undergraduate science and engineering education, noting that improvements in student motivation were often attributed to the novelty of V-Labs rather than their underlying design [31]. Similarly, Sellberg highlighted a significant gap in the literature, pointing to a lack of descriptive, qualitative studies that investigate the everyday instructional practices of virtual laboratories in authentic STEM education contexts [25]. This underscores the need for further research to explore the practical application of virtual laboratories in real-world instructional settings, which would not only enhance their effectiveness but also contribute to advancing theoretical understanding in the field.

Virtual laboratory platforms, such as QLab, have been developed and commercialized to address the pedagogical challenges associated with teaching quantum information science in higher education. However, the effectiveness of these platforms in enhancing teaching and learning outcomes, particularly within the context of general education curricula, remains underexplored. This study aims to investigate the impact of virtual laboratories on the teaching and learning of quantum information science in general education courses in this work.

## III. Methods

To address the challenges in teaching and learning curriculum on quantum information science within general education, a Bell test experiment similar to that used by Lahoz Sanz [22] was developed. Unlike the physical experimental setup described in reference [25], our experiment is conducted on QLab, a commercially available virtual laboratory platform, designed to accommodate undergraduates from both science and arts backgrounds who lack experimental expertise in quantum physics.

**Objective of the research**

1. Determine how the virtual laboratory can affect the learning of students on quantum entanglement.
2. Determine how the virtual laboratory can affect the learning interest of students.

**Research hypotheses**

$H0_1$: Virtual laboratory has no significant impact on the learning outcome.

$H0_2$: Over 80% students benefit from the virtual laboratory.

**Methodology**

The experiment is conducted during a two-hour tutorial session. scheduled after the completion of lectures on quantum entanglement and the Bell test, as illustrated in Fig. 1.

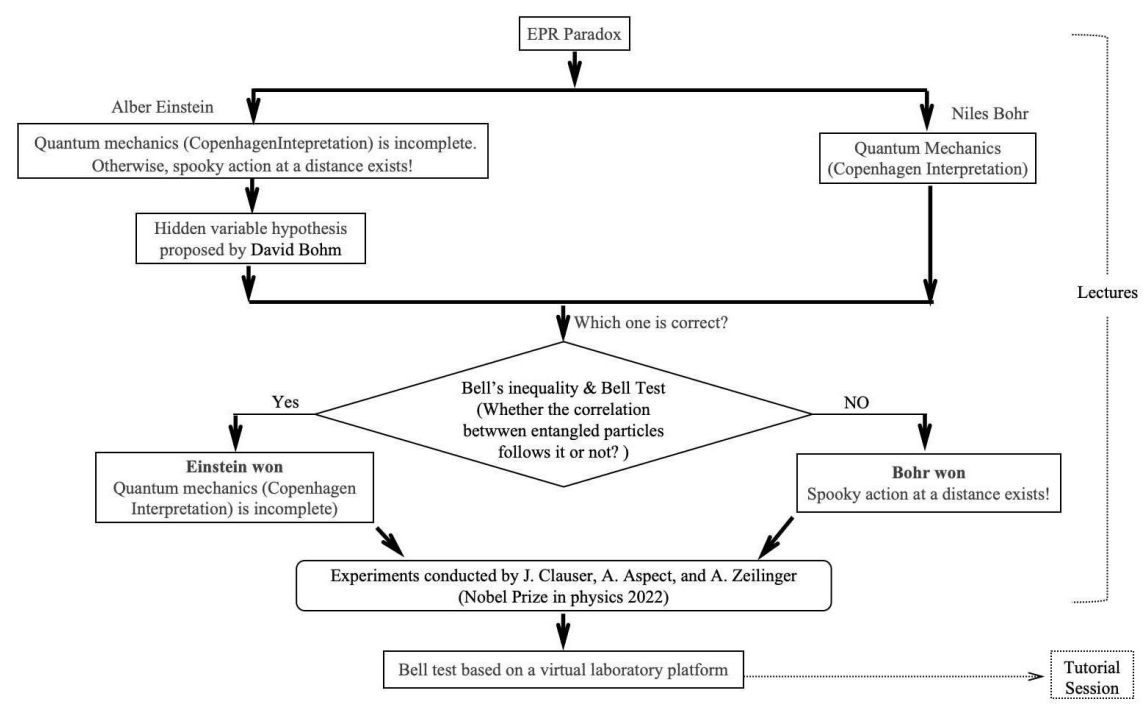

Fig.1 Link between the virtual laboratory and previous lectures

To facilitate effective learning, the following questions are designed and asked at the beginning of the tutorial session:

QA: What is the purpose of Bell test?
QB: What is an entangled state?
QC: How to conduct the Bell test in theory?
During the initial two rounds of virtual laboratory sessions conducted in the academic years 2022-2023 and 2023-2024, these questions were addressed collectively by the entire class. However, in the academic year 2024-2025, the questions were discussed within smaller groups to promote a more collaborative and interactive learning environment.

The interface of the virtual platform for the Bell test is shown in Fig. 2. Similar to the physical setup described in [25], the experimental setup consists of three units, as illustrated in Fig. 3, the generation of a quantum entangled state, the distribution of entangled photons, and the measurement of quantum correlations. A guidebook outlining the principles and operational procedures of the virtual experiment, similar to that in [22], was distributed to students in advance. The fidelity of the platform was ensured through repeated testing prior to implementation, guaranteeing reliable results when correct procedures were followed. Due to the extensive use of optical devices, the instructor provides a detailed explanation of each device's function in theory before starting the experiment.

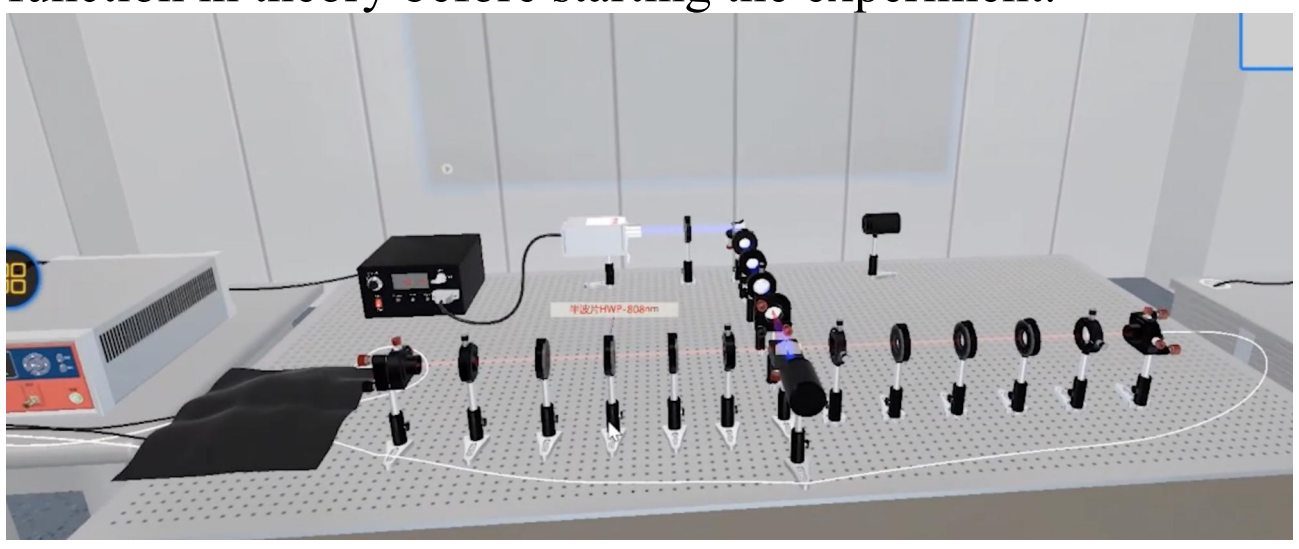

Fig. 2 Virtual experimental setup for the Bell Test in the virtual laboratory platform

In the academic years 2022–2023 and 2023–2024, students conducted the experiments individually. Based on an analysis of student feedback, the instructional approach was revised to encourage collaborative exploration of each device's functions in groups. This modification aimed to promote active learning and enhance student engagement. The instructor provided support whenever students encountered questions about the operation or observed confusing phenomena.

To evaluate the impact of the virtual laboratory on students' learning and engagement, a mixed-methods approach was adopted. Quantitative data collection, drawing on the methodology outlined by Ogbuanya [29], was integrated with qualitative analysis to ensure a comprehensive evaluation.

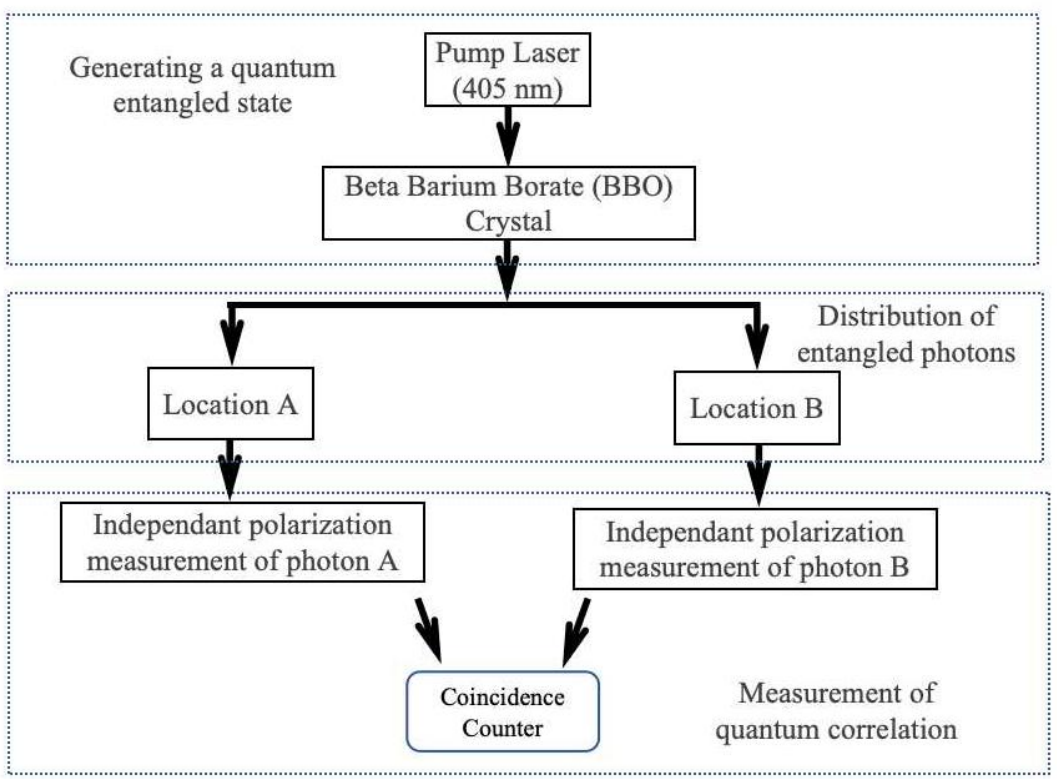

Fig. 3 Schematic diagram of three units in the virtual platform

## IV. Result and Discussion

Following the completion of the virtual experiment, both quantitative and qualitative data were collected from student enrolled through structured surveys administered over three consecutive academic years, 2022–2023, 2023–2024, and 2024–2025. While participation in the surveys was encouraged, it remained entirely voluntary.

Table 1. Questions designed for quantative and qualitative study

| Questions |
| --- |
| Q1: How do you perceive the lab work in this course?<br>Q2: Do you have any additional comments or feedback about the lab work? |

To assess the effectiveness of the virtual laboratory, two types of survey questions were developed, as detailed in Table 1. The first question (Q1) was designed to quantitatively evaluate students' perceptions of the virtual laboratory, while the second question (Q2) was an open-ended prompt intended to gather qualitative insights, allowing students to articulate their opinions and suggestions in an unrestricted manner. As elective general education courses, the enrollment quota ranging from 50 to 75 students per semester. The survey response rates varied across the three academic years, with 52 out of 60 (87%) responses collected in 2022–2023 , 14 out of 72 (19%) in 2023–2024, and 16 out of 49 (33%) in 2024–2025.

- **Results on the first objective of the research**

A. Quantitative analysis on the evaluation of the virtual laboratory

In the initial two semesters, multiple-choice questionnaires were designed for Q1 shown in table 1 to assess students' feedback on laboratory activities. In the third semester, a single-choice survey format was implemented to enhance the clarity and decisiveness of responses. The original five options, (a) 'More lab work activities would be appreciated,' (b) 'Lab work enhanced my interest in optical experiments,' (c) 'Lab work deepened my understanding of the lectures,' (d) 'The laboratory time is limited and would benefit from being extended,' and (e) 'It is too time-consuming and not very rewarding', were consolidated and restructured into four more targeted options: (a) 'It is too time-consuming and not very rewarding,' (b)'Laboratory work deepened my understanding of the lectures and enhanced my interest in optical experiments. More lab activities would be appreciated,' (c) 'The laboratory time is too limited and would benefit from being extended,' and (d) 'Others.' This modification aimed to align the survey format with the objective of obtaining more focused and actionable insights from students, as illustrated in Fig. 4 (a) and 4 (b). An analysis of student feedback revealed that the majority of participants found the virtual laboratory to be a valuable learning tool for depeening their understanding of the lectures, with satisfaction rates of 50%, 85.71%, and 81.25% reported in the semesters of 2022–2023, 2023–2024, and 2024–2025, respectively. This result indicates the effectiveness of virtual laboratory in providing an impactful experiential learning experience.

■ **Chi-square test for the survey results**

We use the Chi-square test to evaluate the validity of the research hypotheses outlined in the previous section.

■ **Testing of $H0_1$**

Table 2. Collected data for Chi-square testing of $H0_1$

| Options for Q1 (Multiple-choice question) | Academic Year 2022-2023 (52 samples in total ) | Academic Year 2023-2024 (14 samples in total ) |
|---|---|---|
| Option1:More lab work activities would be appreciated | 19 (36.54%) | 8 (57.14%) |
| Option2: Lab work enhanced my interest in optical experiments. | 24 (46.15%) | 4 (46.15%) |
| Option3: Lab work deepened my understanding of the lectures. | 26 (50.00%) | 12 (85.71%) |
| Option4 : The laboratory time was limited and it would be beneficial to extend it. | 6 (11.54%) | 2 (14.29%) |
| Option5: It is too time consuming and not very rewarding. | 8 (15.38%) | 0 (0.00%) |

Students are expected to evenly distribute their selections among the five options in the questionnaires as shown in Fig. 4 (a) for both the academic years 2022–2023 and 2023–2024. Accordingly, the expected values are 10.4 and 2.8 for the 2022–2023 and 2023–2024 semesters, respectively. Based on the collected data shown in Table 2, the corresponding Chi-square values are calculated as $\chi^2_{2022\text{-}2023}$=50.71 and $\chi^2_{2023\text{-}2024}$=43.43, both of which exceed the critical Chi-square value of 9.488 at a significance level of $\alpha$ = 0.05 with 4 degrees of freedom. Therefore, the statistical results reject the null hypothesis ($H0_1$), verifying that the virtual laboratory has a significant impact on students' learning.

■ **Testing of $H0_2$**

To test $H0_2$, we refined the survey for the academic year 2024–2025 to include four single-choice questions, as shown in Fig. 4 (b). Students were hypothesized to select the four options (A, B, C, and D) with a distribution of 80.00%, 10.00%, 5.00%, and 5.00%, respectively, as presented in Table 3 and Fig. 4(b). The corresponding expected value of 4 and Chi-square value of $\chi^2_{2024\text{-}2025}$=0.0840 were calculated. The results fail to reject $H0_2$, as $\chi^2$ (2024–2025) is less than the critical Chi-square value of 7.815 at a significance level of $\alpha$ = 0.05 and 3 degrees of freedom. The results indicate that over 80% students benefit from the virtual laboratory, highlighting its significant impact an effective method of experiential learning.

■ **Results on the second objective of the research**

As shown in Fig. 4(a), 46.15% of students in the academic year 2022-2023 and 28.57% in 2023-2024 expressed a preference for additional laboratory activities, particularly those involving hands-on, real-world experiments. This observation aligns with the findings of Ogbuanya, Makransky, and Petersen [29, 34], which underscore the effectiveness of virtual learning environments in fostering student engagement and enhancing their interest in learning. These results directly address the second research objective of the study.

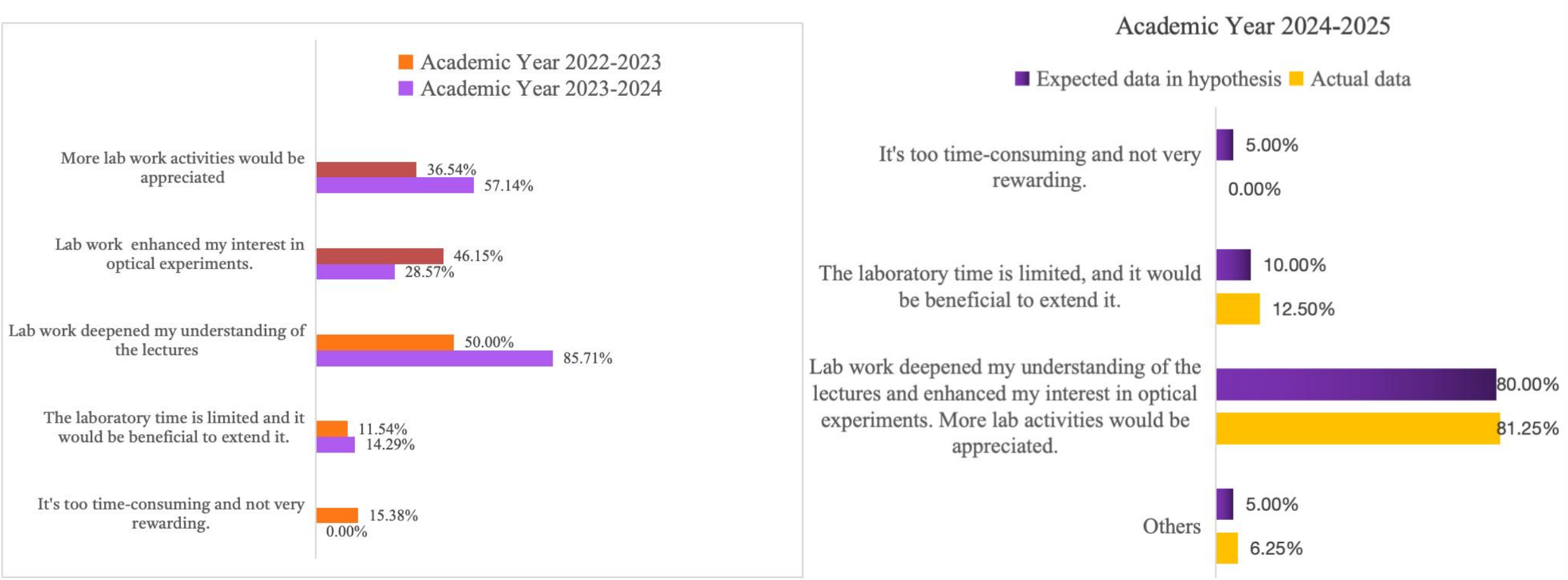

(a) (b)

Fig. 4 Feedback from students of (a) academic years 2022-2023 and 2023-2024 (b) academic year 2024-2025

Table 3. Parameters for Chi-square testing of $H0_2$

| Options for Q1 (Single-choice question) | Academic year 2024-2025 (16 samples in total ) | |
|---|---|---|
| | Actual data | Expected data in hypothesis |
| Option A: Lab work deepened my understanding of the lectures. | 13 (81.25%) | 12.8 (80.00%) |
| Option B: The laboratory time was limited and it would be beneficial to extend it. | 2 (12.5) | 1.6 (10.00%) |
| Option C: It is too time consuming and not very rewarding. | 0 (0.00%) | 0.8 (5.00%) |
| Option D: Others | 1 (6.25%) | 0.8 (5.00%) |

Furthermore, 15.38% of students reported that the virtual experiment was overly time-consuming and not particularly rewarding in the academic year 2022–2023, as illustrated in Fig. 4(a). This indicates that approximately 84.62% of students were satisfied with the virtual laboratory design. In the subsequent two academic years, no students reported dissatisfaction, as shown in Fig. 4 (a) and Fig. 4 (b). These results suggest a progressive improvement in student perceptions of the virtual laboratory over time, further validating its effectiveness as a pedagogical tool.

Table 4. Students' feedback collected in 2022-2023

| Category | | Feedback from students | | |
|---|---|---|---|---|
| | | 2022-2023 | 2023-2024 | 2024-2025 |
| Positive learning outcome | Enhanced Engagement | 1. It was interesting to experience the virtual laboratory for the first time. It would be even better if we could design our own experiments.<br>2. Very interesting<br>3. Lab work is simple and easy to understand.<br>4. Very interesting<br>5. Very good.<br>6. It is okay<br>7. There seems to be a gap between theoretical learning and lab work. At first, facing so many instruments felt a bit overwhelming, but overall, lab work has been a very meaningful experience.* | 1. Very good<br>2. Quite good<br>3. It's very interesting, but the lab software could be a bit more realistic.*<br>4. Lab work is quite interesting, and even though the instruments are simulated, I still find it fascinating. However, I feel that lab work doesn't significantly help in understanding the course content. | 1. It is very interesting and I like it very much.<br>2. Very Good<br>3. A lot of fun when conducting lab work.<br>4. Quite good<br>5. It is good<br>6. Very interesting |
| | Desire for More Virtual and/or Physical Labs | 8. Having hands-on physical experiments would be even better.<br>9. Having a real lab would make it even more engaging.<br>10. Besides virtual laboratory, it is better to add some physical lab if possible.<br>11. Lab sessions should be included in every class.<br>12. It would be even better if there were physical experimental equipment<br>13. After completing it, I felt bland. There should be more interactive and hands-on components.<br>14. Looking forward to more lab works<br>15. If possible, I hope to increase the number of lab works and have the opportunity to physically interact with the experimental instruments.<br>16. The weight of lab work could be increased, and students could be encouraged to personally perform the related mathematical derivations and explore concepts hands-on to deepen their understanding (rather than solely relying on a virtual lab)<br>17. The lab work duration is too short; I suggest extending the allocated time. It has increased my interest in optical experiments, and adding more lab work content would further enhance the experience.<br>18. The efficiency is a bit low, and the content seems somewhat limited.<br>19. Looking forward to more lab works<br>20. Simulated experiments might lack a certain degree of credibility. | 5. I really love lab work and would love to have more of it!<br>6. It would be great to use real equipment for demonstrations. | 7. I hope to have more practical labs instead of virtual labs.<br>8. If possible, I would like to have the opportunity to conduct experiments in the laboratory.<br>9. It is better to have one or two virtual labs since too many sessions would be too stressful. |
| | Improved learning | 21. Not bad. It feels more efficient than just listening to lectures.<br>22. There seems to be a gap between theoretical learning and lab work. At first, facing so many instruments felt a bit overwhelming, but overall, lab work has been a very meaningful experience.<br>23. Related course content became more intuitive, which broadened my horizons. | 7. Lab work has enhanced my grasp of the course content's finer details. I often realize during hands-on practice that some knowledge points weren't well understood. Through the experiments, I've been able to solidify my understanding and enrich the course content overall. | |
| | Accessibility and safety | 24. It is friendly for non-experimental students, as it eliminates issues caused by operational errors and ensures safety in experiments. However, it might lack a sense of realism. Since it is expensive, though real equipment is even costlier, I hope it can be more beneficial for physics majors. | | |
| Challenges and barriers | Understanding Issues | 1. Don't understand the underlying principles.<br>2. I completed it but still found it difficult to understand. | | |

| Category | | Feedback from students | | |
|---|---|---|---|---|
| | | 2022-2023 | 2023-2024 | 2024-2025 |
| | | 3. There seems to be a gap between theoretical learning and lab work. At first, facing so many instruments felt a bit overwhelming, but overall, lab work has been a very meaningful experience.*<br>4. Since the experiments are not exactly the same as those in the main course, it can be challenging to fully understand them.<br>5. Sometimes, just inputting data based on the experiment manual doesn't lead to a good understanding—it feels like mere mechanical input.<br>6. Useless<br>7. It is better to deliver lectures by the instructor.<br>8. It feels like I just collect some data.<br>9. I can't rule out that arriving late might have contributed, but I feel I lack a clear, intuitive understanding of the experiment as a whole. Although I managed to measure the data, I don't have a clear grasp of the overall purpose of the experimental design. #<br>10. I don't really know how to do it.<br>11. Personally, I feel the efficiency is a bit low, probably because we can only operate on the computer. It seems there might be an issue with the software? For example, even though my instrument parameters were the same as other students', the results turned out differently. # | 1. I feel like I haven't fully grasped the principles behind the experiments.<br>2. Lab work is quite interesting, and even though the instruments are simulated, I still find it fascinating. However, I feel that lab work doesn't significantly help in understanding the course content. * | |
| Suggestions for Improvement | Increased interactivity and self-exploration | 1. It would be helpful to explain the experimental process and concepts further during the lectures.<br>2. The teacher is amazing and guided us step by step on how to conduct the experiment. However, if we were given the opportunity to explore on our own first, we might have gained even more valuable insights<br>3. The content seems to be a bit limited. It often feels like we're just following the teacher's instructions to input and record data, without much room for independent exploration or expansion. | 1. It would be great to include more opportunities for self-exploration.<br>2. The software on some computers in the computer lab runs very slowly, and switching computers can waste time. If classmates don't follow the teacher's instructions carefully, it also causes delays. For those who finish calculating the data early, they often have nothing to do afterward. It would be helpful to either add more tasks or encourage students to assist each other in groups.<br>3. It would be best to have complete video tutorials. Relying solely on PDF files makes it difficult to fully understand the material in advance. This lack of understanding often leads to anxiety during the lab sessions, as students worry about not completing the experiment on time. As a result, the primary goal during the lab shifts to simply finishing the experiment, which prevents students from fully thinking through and connecting the experiment to the classroom knowledge. Additionally, learning the details of the experiment in advance would reduce the anxiety of students getting stuck at certain steps and lessen the burden on teachers having to answer repeated questions. | |
| | Better software functionality | 4. The software has bugs, the operation feels a bit clunky, and, after all, it's just software.<br>5. A feature to save complete sets of instrument value settings could be added.<br>6. The software's documentation is malfunctioning and cannot be opened.<br>7. The course manual is very detailed, but the software isn't particularly user-friendly and has poor operability.<br>8. The software interface is somewhat difficult to operate. | | |

*: Comments can be classified into two categories with unrelated parts in gray.
#: With possible special reason.

## B. Qualitative analysis on the evaluation of the virtual laboratory

An open-ended question, Q2 shown in Table 1, was designed to gather students' feedback on the virtual laboratory. Responses were collected from 41 out of 52 (78.85 %), 11 out of 14 (78.57%), and 9 out of 16 (56.25%) students who completed the questionnaire during the academic years 2022–2023, 2023–2024, and 2024–2025, respectively. To analyze the data, thematic coding was employed to categorize the feedback into three key themes, "Positive Learning Outcomes," "Challenges and Barriers," and "Suggestions for Improvement," as summarized in Table 4. The results reveal that 24 of 41 (58.54%), 7 out of 11 (63.64%), and 9 out of 9 (100.00%) students across the three academic years reported that the virtual laboratory enhanced their engagement, improved their learning outcomes, and provided a safe and supportive environment for experiential learning. In particular 13 of 41 (31.71%), 2 of 11 (18.18%) and 3 of 9 (33.33%) students expressed a strong interest in expanding laboratory opportunities, including both virtual and physical formats. Notably, these students were not majoring in physics or quantum information science, which highlights the interdisciplinary appeal and accessibility of the virtual laboratory.

These findings underscore the value of the virtual laboratory in motivating undergraduate students to explore quantum physics, aligning with the reports in [29, 34]. They also highlight its significant role in fostering student engagement and interest in physics, as supported by the previous study [31], particularly among learners from diverse educational backgrounds. Furthermore, the increasing proportion of students reporting positive learning outcomes over the three academic years suggests that the virtual laboratory is an effective tool for enhancing experiential learning. It demonstrates the potential to address the varied needs of students, regardless of their academic specialization.

Moreover, during the first two academic years, 2022–2023 and 2023–2024, 3 out of 41 (7.32%) and 3 out of 11 (27.27%) students, respectively, provided suggestions for improving the teaching approach by increasing interactivity and self-exploration. These suggestions formed a valuable foundation for refining the teaching methodology in subsequent academic years. Specifically, group discussions on QA, QB, and QC, as outlined in Part III, were integrated into the third round of virtual laboratory sessions during the academic year 2024–2025. These sessions encouraged students to collaborate in groups to conduct experiments, fostering teamwork and active engagement.

As shown in Table 4, 11 out of 41 (8.94%) and 2 out of 11 (18.18%) students reported challenges related to "Understanding Issues" during lab work in the first two academic years. These findings underline the need for stronger integration of theoretical concepts with practical tasks in the virtual laboratory, despite some cases being attributed to late arrivals and insufficient engagement in lectures.

To address these issues, targeted improvements were implemented based on student feedback. These included incorporating collaborative group discussions, distributing lab reading materials in advance to allow students sufficient time to prepare, and providing clearer guidance during experiment. By the academic year 2024–2025, no students reported such issues, demonstrating the effectiveness of these adjustments. The results in Fig. 4(b) further support these findings, as no students identified the virtual laboratory as being too time-consuming or unrewarding. These outcomes indicate that aligning the virtual laboratory activities more closely with students' needs and expectations not only enhances comprehension but also increases engagement and the perceived value of the lab sessions. This iterative approach underscores the importance of proactively using student feedback to refine teaching strategies and ensure that virtual learning tools remain practical, effective, and responsive to learner challenges.

Moreover, 5 out of 41 (12.20%) students provided feedback on the usability and realism of the virtual laboratory platform. While feedback on usability was no longer observed in the academic years 2023–2024 and 2024–2025 following the platform update, the comments on realism highlight students' interest in physical laboratory experiences. However, implementing physical labs in general education courses is often impractical due to resource constraints and the specialized expertise required for quantum optical experiments. This underscores the potential of hybrid experiential learning as a viable approach to bridge the gap between theoretical concepts and practical applications in STEM education. Such an approach could enhance both teaching and learning outcomes while presenting a valuable opportunity for exploration and innovation in pedagogical practices.

# V. CONCLUSION

This study assessed the pedagogical effectiveness of a virtual laboratory platform, QLab, in general education courses on quantum information science. Designed to accommodate students from diverse academic backgrounds, QLab demonstrated its capacity to overcome traditional barriers to learning abstract and counterintuitive quantum concepts. Over three consecutive academic years (2022–2023, 2023–2024, and 2024–2025), qualitative and quantitative analyses revealed that over 80% of participants reported significant learning benefits. These findings underscore QLab's ability to enhance conceptual understanding and promote the practical application of quantum phenomena across disciplines.

The results consistently highlight QLab's role as a transformative educational tool. By addressing inherent challenges in teaching quantum information science, the platform fosters deeper engagement and accessibility, even for students lacking advanced mathematical or technical expertise. Furthermore, QLab eliminates safety risks associated with high-power lasers in traditional quantum optics laboratories and avoids potential health concerns linked to prolonged use of immersive virtual reality devices. These advantages position QLab as an effective, safer, and more inclusive alternative for both physics majors and non-specialists.

While the study provides strong evidence of QLab's pedagogical value, certain limitations warrant consideration. The response rate, which was high (87%) during the first academic year (2022–2023), declined in subsequent years, potentially affecting the robustness of longitudinal conclusions. Nevertheless, the large initial sample size and the use of triangulated qualitative and quantitative methods ensure a reliable foundation for analysis. Future research should focus on increasing response rates and expanding the sample size to strengthen the validity of findings and further explore the platform's broader applicability.

In conclusion, QLab exemplifies the potential of virtual laboratory platforms to transform STEM education. By bridging gaps in accessibility and engagement, it promotes inclusivity and scalability, expanding participation in specialized topics such as quantum information science. These findings underscore the importance of integrating innovative pedagogical tools into diverse educational contexts, paving the way for broader adoption of advanced virtual learning methodologies. As such, QLab serves as a model for advancing STEM education and fostering interdisciplinary learning in the 21st century.

## Conflict of Interest

The authors declare no conflict of interest.

## Funding

This research was funded by the Teaching Innovation Grant 2023 at the Chinese University of Hong Kong, Shenzhen and Teaching Reform Project of Guangdong.